\documentclass[fleqn,10pt]{wlscirep}
\usepackage[utf8]{inputenc}
\usepackage[T1]{fontenc}
\usepackage{hyperref}
\usepackage[final]{pdfpages}
\makeatletter
\AtBeginDocument{\let\LS@rot\@undefined}
\makeatother
\hypersetup{
    colorlinks = false,
    linkbordercolor = {white},
}

\usepackage{amsmath,amssymb,booktabs,latexsym,MnSymbol,bm}

\newcommand\ie{\mathrel{{\leftrightarrow}}}
\newcommand*\IPpp{\mbox{$I\!P\!\!+\!\!+$}}
\newcommand*\Ipp{\mbox{$I\!\!+\!\!+$}}
\newcommand*\Ppp{\mbox{$P\!\!+\!\!+$}}
\newcommand*\IpPp{\mbox{$I\!\!+\!\!P+$}}
\newcommand*\IpPm{\mbox{$I\!\!+\!\!P-$}}
\newcommand*\ImPp{\mbox{$I\!\!-\!\!P+$}}
\newcommand*\ImPm{\mbox{$I\!\!-\!\!P-$}}

\title{Gender shapes the relationship between productivity and journal prestige in science}

\author[1]{\normalsize V{\'i}tor H. Ribeiro} 
\author[1]{\normalsize Andre S. Sunahara} 

\author[2]{\normalsize Golnaz~Shahtahmassebi} 

\author[3,4,5,6,*]{\normalsize Matja{\v z}~Perc} 

\author[1,*]{\normalsize Haroldo V. Ribeiro} 

\affil[1]{\normalsize Departamento de F\'isica, Universidade Estadual de Maring\'a, Maring\'a, PR 87020-900, Brazil}

\affil[2]{\normalsize School of Science and Technology, Nottingham Trent University, Clifton Lane, Nottingham, United Kingdom}

\affil[3]{\normalsize Faculty of Natural Sciences and Mathematics, University of Maribor, Koro{\v s}ka cesta 160, 2000 Maribor, Slovenia}
\affil[4]{\normalsize Community Healthcare Center Dr. Adolf Drolc Maribor, Vo{\v s}njakova ulica 2, 2000 Maribor, Slovenia}
\affil[5]{\normalsize Department of Physics, Kyung Hee University, 26 Kyungheedae-ro, Dongdaemun-gu, Seoul 02447, Republic of Korea}
\affil[6]{\normalsize University College, Korea University, 145 Anam-ro, Seongbuk-gu, Seoul 02841, Republic of Korea}

\affil[*]{\footnotesize email: matjaz.perc@gmail.com; hvribeiro@uem.br}

\begin{abstract}
Gender disparities in academia manifest and persist in various aspects of the scientific enterprise, yet their influence on the interplay between research productivity and journal prestige remains underexplored. Here we analyze the academic trajectories of over 6,000 elite Brazilian researchers by jointly tracking their annual productivity and the average prestige of the journals in which they publish. By projecting individual career years onto a standardized productivity-prestige plane and applying Bayesian hierarchical modeling, we find that male researchers are more likely to follow productivity-oriented trajectories and are markedly overrepresented in the hyperprolific region of this plane. Female peers, in contrast, more often occupy regions that prioritize journal prestige over publication quantity. Although male researchers publish more throughout their careers, their female counterparts achieve comparable or higher average journal prestige, particularly in later career stages and among outlier individuals. Male researchers also exhibit greater temporal persistence in their productivity and impact levels and are especially averse to simultaneously changing both metrics compared to their female peers. Among non-outliers, productivity and career age have a negative overall impact on the average journal prestige of researchers of both genders, with slightly stronger effects observed among female researchers; however, these patterns vary across disciplines, highlighting the complexity and heterogeneity of academic careers.
\end{abstract}

\begin{document}

\flushbottom
\maketitle
\thispagestyle{empty}

\section*{Introduction}

Gender disparities remain prevalent across several domains, reflecting persistent inequalities shaped by historical biases and cultural norms that continue to influence key aspects of life, including economics~\cite{blau2017wage, kim2024genderpaygap}, health~\cite{tannenbaum2019sex, shansky2021biological}, crime~\cite{kruttschnitt2013gender, pessa2025structural}, and literature~\cite{schulz2019literature, stuhler2024gender}. In academia, these disparities are particularly multifaceted, manifesting across collaboration patterns~\cite{zeng2016collaboration, uhly2017patterns}, academic mobility~\cite{zhao2021international, zhao2023migration}, funding opportunities~\cite{ley2008gender, duch2012possible}, self-promotion of research~\cite{peng2025gender}, participation in scientific events~\cite{anteneodo2023physics}, peer evaluation processes~\cite{robyn2009name, liu2023gender} and also in the development of novel ideas and the ability to produce disruptive or interdisciplinary science~\cite{liu2023interdisciplinary, zhang2024composition, liu2024novelty}. Beyond their societal implications, efforts to foster diversity within the scientific workforce are crucial for addressing global challenges and promoting economic development, as diversity has been shown to enhance creativity and accelerate scientific and technological discoveries, among other relevant benefits~\cite{xie1998sex, page2008difference, nielsen2018making, alshebli2018preeminence, bell2019becomes, hofstra2020diversity, koning2021we, yang2022gender}.

The persistence of gender inequalities in scientific careers is particularly evident in research productivity and scientific impact, often measured by publication counts and citation-based metrics. Gender differences in these indicators have a long and well-documented history, exemplified by the Matthew Effect, which describes how early advantages -- such as initial recognition, access to resources, or prestigious affiliations -- accumulate over time, disproportionately benefiting established scholars~\cite{merton1968matthew}; and the Matilda Effect, which captures how female scientists' contributions have historically been systematically overlooked, undervalued, or even attributed to male colleagues, leading to persistent under-recognition of their work despite equivalent accomplishments~\cite{rossiter1993matilda, lincoln2012matilda}. Recent studies show that diversity in the scientific workforce is unevenly increasing and at slow rates~\cite{fry2021stem, boekhout2021gender, wef2025global}, suggesting that both effects continue to contribute to gender-based disparities in academic careers. Moreover, emerging evidence suggests that these disparities may be further exacerbated in the era of artificial intelligence, as male researchers are more likely than their female peers to use generative tools and experience a greater increase in productivity~\cite{tang2025IA}.

This concern is particularly pressing given the growing reliance of research evaluations on bibliometric assessments~\cite{cameron2005trends, traag2019systematic, meirmans2019science, mckiernan2019meta}. While recent scholarship places greater emphasis on the quality and innovativeness of scientific outputs~\cite{liu2024novelty, liu2023interdisciplinary}, traditional metrics of productivity and impact -- despite facing considerable criticism~\cite{dora2012, nuffield2014, hicks2015bibliometrics, wilsdon2016metric, lariviere2019jifhistory} -- remain widely used and continue to influence recognition within the scientific community. In this context, the relationship between research productivity and impact has, however, proven to be complex and influenced by many factors, including disciplinary differences, career stage, scale effects, individual variability, and the presence of outlier individuals~\cite{white1978relation, feist1997quantity, haslam2010quality, bosquet2013academics, abramo2014authors, sandstrom2016quantity, lariviere2016many, michalska2017and, kolesnikov2018researchers, bornmann2019productivity, forthmann2020investigating, sunahara2021association}. Despite increasing awareness of these issues, many studies still overlook these complexities, often relying only on aggregate metrics and ignoring the heterogeneity of academic trajectories. Additionally, while factors such as collaboration, mobility, and funding have received significant attention, the potential role of gender in the association between productivity and impact remains particularly underexplored, obscuring how systemic quantitative pressures may differentially constrain the capacity for risk-taking and high-quality innovation among female researchers. Here we address this gap by analyzing the careers of over six thousand female and male scholars recognized as part of the Brazilian scientific elite across multiple disciplines. Unlike many academic systems in the Global North, in Brazil, the vast majority of high-impact research is conducted within public (federal and state) universities. Admission to these institutions is strictly regulated through public civil-service examinations, which grant tenure early in researchers' careers, but also require researchers to divide their time between comparable substantial teaching duties -- often at both undergraduate and graduate levels -- and scientific investigation. Furthermore, research funding and evaluation are centralized under federal agencies, primarily the National Council for Scientific and Technological Development (CNPq) and the Coordination for the Improvement of Higher Education Personnel (CAPES). Together, these features confer distinctive characteristics to the Brazilian research ecosystem, making it a particularly informative case for investigating how productivity relates to impact, and how this relationship varies by gender across disciplines. Based on this scientific ecosystem, we reconstruct each researcher's trajectory by jointly tracking the yearly evolution of their productivity and the average prestige of the journals in which they publish -- quantified using journal impact factor and SCImago journal rank -- and then apply a recently introduced standardization approach~\cite{sunahara2021association} that accounts for the presence of outlier individuals, discipline-specific publication practices, inflation-like effects, and size-dependent biases in the average journal prestige evaluation~\cite{antonoyiannakis2018impact, antonoyiannakis2020impact}, to probe gendered patterns in the relationship between productivity and journal prestige. 

Our findings reveal that male researchers tend to adopt productivity-driven strategies, whereas female researchers more often pursue impact-oriented trajectories, revealing a distinct gendered divergence in the quantity-versus-quality trade-off. These divergent strategies lead to consistently higher productivity among male scholars across career stages; nevertheless, despite publishing at lower rates, female researchers tend to achieve comparable -- or even higher -- average journal prestige, particularly in the later stages of their careers and among outlier individuals (those whose productivity and/or average journal prestige markedly exceed the typical values of their disciplines). Compared to females, male researchers are less likely to shift their productivity or journal impact levels, especially among outlier individuals, and achieving extremely high productivity reduces the likelihood of reaching outlier performance in average journal prestige for female scholars more than for males. In general, productivity and career age are negatively associated with average journal prestige for both genders, with slightly stronger effects observed among female scholars, despite substantial gender-specific differences across disciplines. These findings further underscore the importance of adopting gender-sensitive approaches in research evaluation to foster a more equitable, fair, and efficient academic environment. They also contribute to the broader literature on gender disparities by offering gendered insights from a major research ecosystem outside North America and Europe -- one that is well-developed, data-rich, and yet underrepresented in global discussions on academic inequalities.

\section*{Results}

\subsection*{Gendered patterns in the occupation of journal prestige versus productivity plane}

\begin{figure*}[ht!]
    \centering
    \includegraphics[width=1\linewidth]{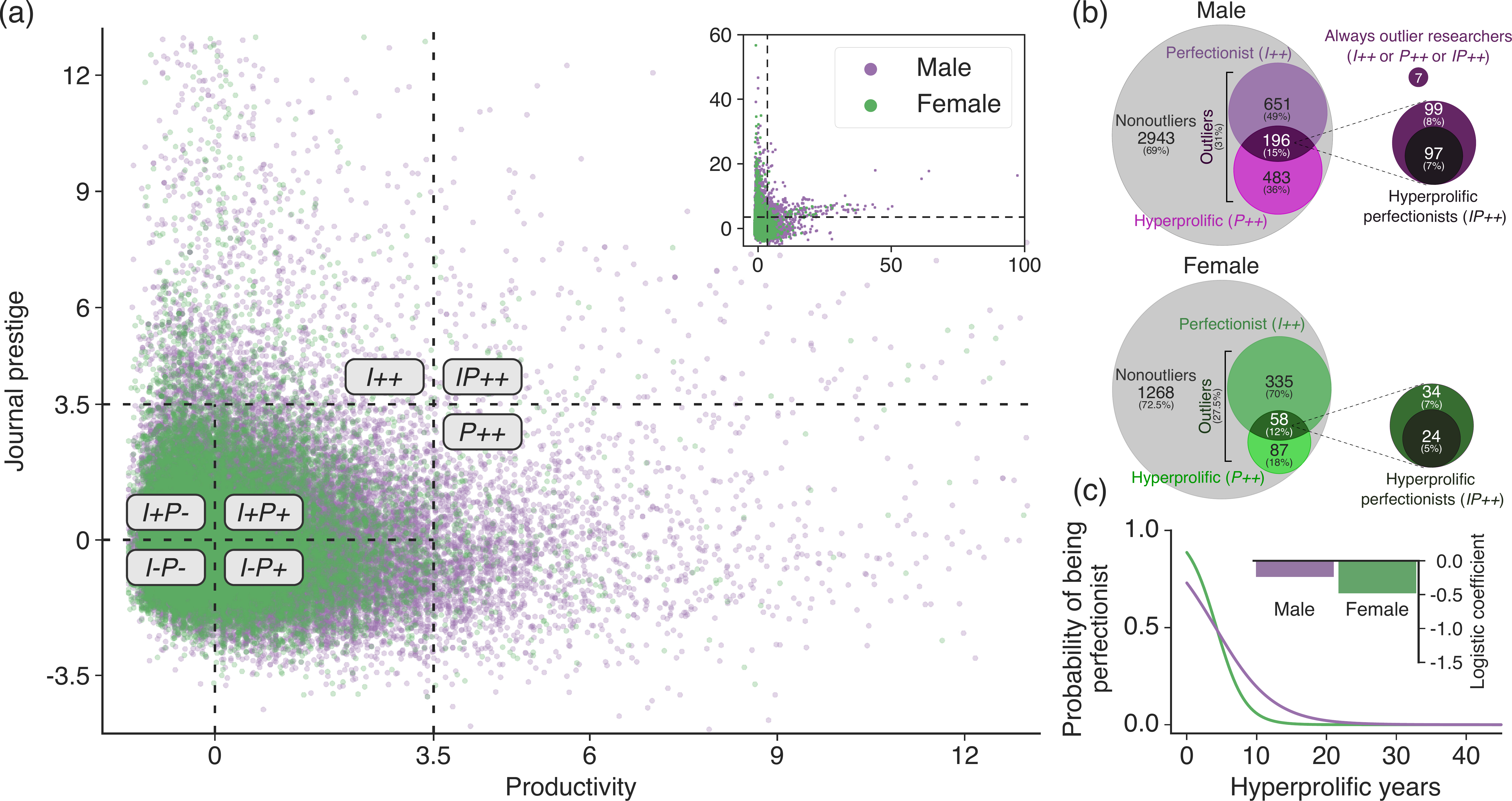}
     \caption{Gendered patterns in the journal prestige-productivity plane. (a) Scatter plot showing the relationship between yearly average journal prestige and productivity across career years of researchers in our dataset. Each point represents a career year of a male (purple) or female (green) researcher, with both variables expressed in standard score units. One unit of productivity corresponds to a one standard deviation above (positive) or below (negative) the average productivity within the researcher's discipline and year. Similarly, one unit of journal prestige represents one standard deviation above (positive) or below (negative) the expected value given the same productivity level within the researcher's discipline and year. The plane is divided into four non-outlier sectors (\IpPp{}, \IpPm{}, \ImPp{}, and \ImPm{}), based on whether metrics are above or below the disciplinary averages, and three outlier sectors (\IPpp{}, \Ppp{}, and \Ipp{}) corresponding to exceptionally high values of journal prestige and/or productivity. The inset depicts the full extent of the plane. (b) Venn diagrams illustrating the set relationships of our classification of researchers into non-outliers (no career year in an outlier sector), perfectionists (at least one career year in \Ipp{}), hyperprolifics (at least one career year in \Ppp{}), and hyperprolific perfectionists (at least one career year in \IPpp{}). The upper diagram displays the results for male researchers, while the lower one shows the results for female researchers. (c) Probability of finding male (purple) and female (green) perfectionist researchers as a function of the number of hyperprolific years in their careers, as estimated via a logistic regression model. The inset shows the fitted logistic coefficients.}
     \label{fig:1}
\end{figure*}

The dataset used in our work is essentially the same as that introduced by Ref.~\cite{sunahara2021association} and comprises the publication records of 6,028 Brazilian researchers (1,748 females) holding a nationally prestigious fellowship (the Research Productivity Fellowship) awarded from the 1970s onward by the National Council for Scientific and Technological Development (CNPq, Conselho Nacional de Desenvolvimento Cient\'ifico e Tecnol\'ogico) to scholars producing high-quality research (see Methods for details). The data were obtained from the national curriculum platform, \textit{Plataforma Lattes}, which has been maintained by the Brazilian government since the late 1990s and is widely used by institutions and funding agencies in the country. Active researchers, particularly fellowship holders, are required to regularly update their publications and research achievements, making the Lattes curriculum the \textit{de facto} source of academic profiles in Brazil. As these curricula are self-reported, our dataset is not affected by the name disambiguation issues that commonly hinder other databases used in science-of-science studies. Although researchers in our dataset represent Brazil's scientific elite, they span diverse disciplines and institutions, providing comprehensive and systematic coverage of the country's scientific enterprise. Furthermore, focusing on this elite group enables the examination of patterns within a relatively homogeneous cohort, allowing us not only to identify gendered patterns persisting at the highest levels of academic recognition but also to explore potentially distinct gender strategies underlying high academic distinction.

We track the joint evolution of each researcher's annual publication count and the average journal prestige of their publications. Journal prestige is estimated using Clarivate's Journal Impact Factor (JIF) and Scopus's SCImago Journal Rank (SJR), and although JIF and SJR differ substantially in their definitions, the results are consistent across both metrics. Therefore, we present the analyses based on JIF in the main text, with corresponding comparisons using SJR reported in the Supplementary Information. It is well established that raw numbers of productivity and average journal impact are affected by inflationary trends and disciplinary differences in publication practices~\cite{sinatra2016quantifying, sunahara2021association}. Additionally, average journal prestige is further subject to a size effect that may bias comparisons among researchers with different productive levels~\cite{antonoyiannakis2018impact, antonoyiannakis2020impact, sunahara2021association}. To account for these biases, we apply a normalization procedure~\cite{sunahara2021association} that defines robust standard scores relative to discipline and year for publication counts, and relative to discipline, year, and productivity level for average journal prestige (see Methods for details). Under this standardization, positive values of normalized productivity indicate researchers with publication counts above the average of their counterparts in the same discipline for a given year, while negative values reflect output below this average. Similarly, positive values of normalized average journal prestige indicate researchers whose average journal prestige exceeds the level expected for their productivity within a given discipline and year, while negative values denote researchers whose journal prestige is below this expectation. 

The combined use of these standardized measures thus allows us to aggregate researchers across disciplines and analyze their performance without bias, focusing on potential gendered patterns in the association between productivity and journal prestige. Figure~\ref{fig:1}(a) depicts a scatter plot where each point represents the normalized values of productivity and average journal prestige for a given career year of a researcher in our dataset, with male career years shown in purple and female career years in green. This representation reveals that researchers' performance on both metrics is highly skewed toward positive values, with some scholars exhibiting exceptionally high productivity and average journal prestige in specific years of their careers. Following Ref.~\cite{sunahara2021association}, we partition this plane into three outlier sectors and four non-outlier sectors based on values of productivity ($P$) and average journal impact ($I$). The outlier sectors \Ipp{} and \Ppp{} correspond to career years marked by exceptionally high average journal impact ($I>3.5$ and $P<3.5$) and productivity ($I<3.5$ and $P>3.5$), respectively, while the sector \IPpp{} comprises career years exhibiting exceptionally high values on both metrics ($I>3.5$ and $P>3.5$). The remaining non-outlier region is further divided into: \IpPp{}, with above-average journal prestige and productivity ($I>0$ and $P>0$); \IpPm{}, with above-average journal prestige but below-average productivity ($I>0$ and $P<0$); \ImPp{}, with below-average journal prestige and above-average productivity ($I<0$ and $P>0$); and \ImPm{}, with below-average performance on both metrics ($I<0$ and $P<0$).

Among female researchers, 72.5\% never assess the outlier sectors, a fraction slightly higher than the 69\% observed among their male counterparts. These individuals are categorized as non-outliers researchers, whereas those with at least one career year in an outlier sector are classified into three categories: perfectionist (at least one career year in sector \Ipp{}), hyperprolific (at least one career year in sector \Ppp{}), and hyperprolific-perfectionist (at least one career year in sector \IPpp{}). Figure~\ref{fig:1}(b) presents Venn diagrams showing the set relationships among these categories, separated by gender. Although the overall proportion of outliers is comparable between male and female researchers (31\% vs. 27.5\%, respectively), their distribution across the three categories differs significantly. A substantial majority -- 70\% -- of female outliers are solely perfectionists, compared with 49\% of male outliers. Conversely, 18\% of female outliers are exclusively hyperprolific, significantly lower than the 36\% observed among male outliers. Additionally, only seven male researchers have had their entire careers, as covered by our dataset, within outlier sectors. Moreover, among outliers, 15\% of males and 12\% of females perform both as perfectionists and hyperprolifics, and only 5\% of females and 8\% of males are hyperprolific-perfectionists. Thus, only a small minority simultaneously achieves outlier status in productivity and journal impact. Indeed, the logistic regression results (see Methods for details) presented in Figure~\ref{fig:1}(c) show that a higher number of hyperprolific career years is associated with a significantly lower likelihood of performing as a perfectionist for both genders; however, the impact is more pronounced among female researchers. For instance, while ten hyperprolific years corresponds to a near-zero probability of a female outlier researcher being a perfectionist, a male outlier, in the same conditions, exhibits approximately an 18\% chance of achieving this status.

\begin{figure*}[ht!]
    \centering
    \includegraphics[width=0.95\linewidth]{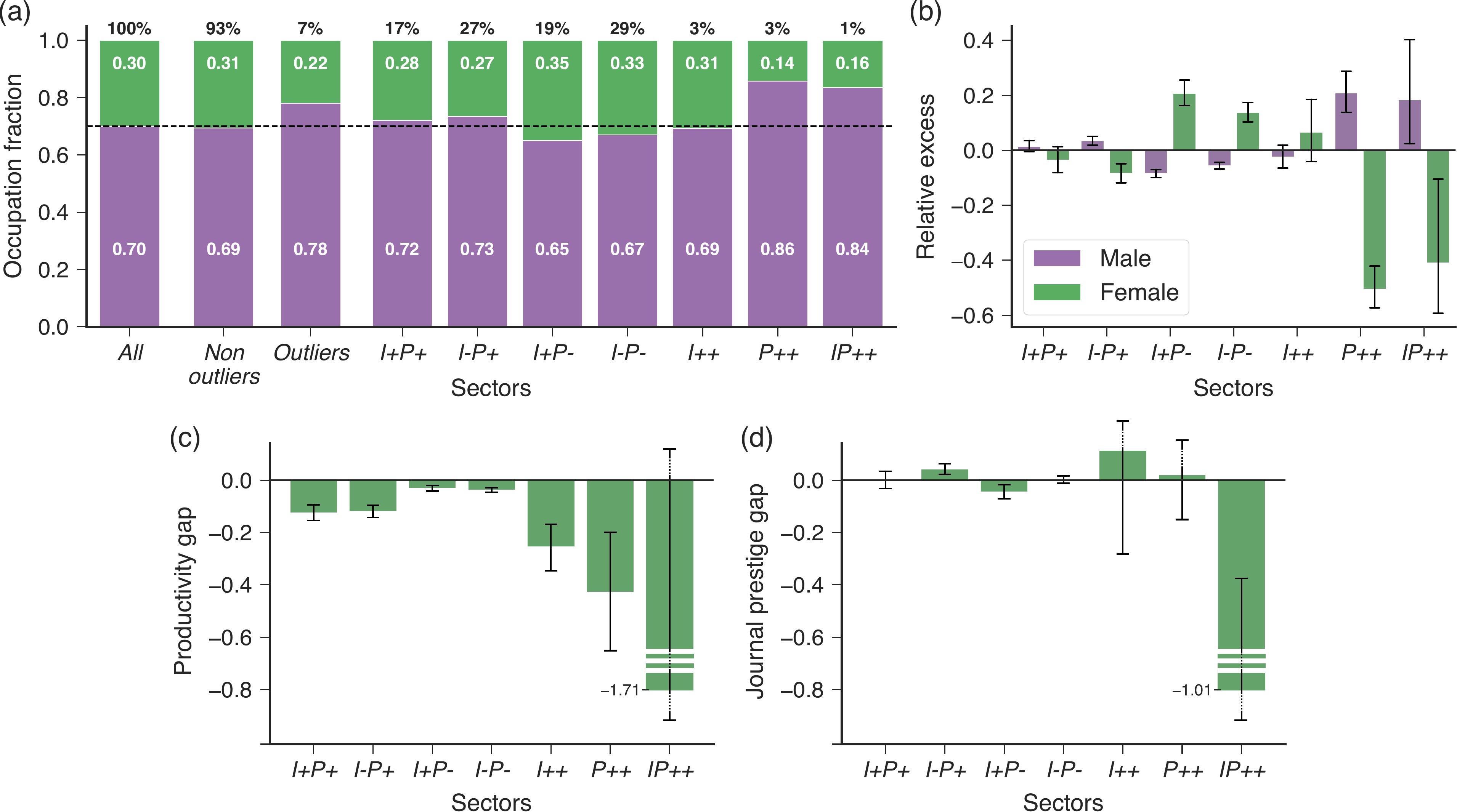}
    \caption{Gendered patterns in the occupation of the journal prestige-productivity plane. (a) Occupation fraction of career years of male (purple) and female (green) researchers in different sectors of the plane, compared with the overall gender prevalence of career years (first bar and dashed line). Numbers above the bars indicate the total percentage of career years per sector, while numbers within the bars denote the gender-specific occupation fractions. (b) Excess of career years of male (purple) and female (green) researchers in each sector relative to a null model in which gender is randomly shuffled across all career years. Error bars indicate 95\% confidence intervals from 1,000 realizations of the null model. (c) Productivity gap, defined as the difference between the average productivity of female and male researchers within each sector. (d) Journal prestige gap, defined as the difference between the average journal prestige of female and male researchers within each sector. Error bars in panels (c) and (d) represent 95\% bootstrap confidence intervals.}
     \label{fig:2}
\end{figure*}

The lower female representation among researchers in our dataset ($28.9\%$) closely aligns with their share of total career years, which accounts for 30\% of the 75,628 career years in our dataset. Assuming, thus, no gendered patterns in the occupation of the average journal prestige-productivity plane, we would expect approximately 30\% of the career years in each sector of this plane to correspond to female researchers. However, the gender differences observed in the classifications of researchers into non-outlier and outlier categories suggest otherwise. Figure~\ref{fig:2}(a) shows the occupation fraction in sectors of the plane for both genders, with reference to the overall representation. While non-outlier years are only slightly overrepresented in female careers, outlier career years are significantly underrepresented, accounting for 22\% of female career years. Given that the proportion of outlier researchers is similar across genders, this finding suggests that male researchers are more likely to access outlier sectors throughout their careers. In the non-outlier sectors, female career years are underrepresented in those prioritizing productivity (\IpPp{} and \ImPp{}) and overrepresented in those not prioritizing productivity (\IpPm{} and \ImPm{}), particularly in \IpPm{}, where female career years constitute 35\% of the total. This pattern becomes more pronounced in the outlier sectors emphasizing extreme productivity, where female career years account for only 14\% in \Ppp{} and 16\% in \IPpp{}, compared to 31\% in the perfectionist sector \Ipp{}.

To statistically substantiate these patterns, we estimate the excess of career years in each sector for both genders relative to a null model in which gender was randomly shuffled across all career years. For each sector and gender, we calculate the relative excess as ${(f-\langle f_{\text{null}}\rangle)}/{\langle f_{\text{null}}\rangle}$, where $f$ is the empirical fraction of years within the sector, and $\langle f_{\text{null}}\rangle$ is average fraction observed in 1,000 realizations of the null model. Figure~\ref{fig:2}(b) shows that the relative excess corroborates the observed trends of sector occupation. Specifically, female researchers exhibit negative relative excess in all productivity-prioritized sectors (\IpPp{}, \ImPp{}, \Ppp{}, and \IPpp{}) and positive relative excess in sectors where productivity is not prioritized (\IpPm{}, \ImPm{}, and \Ipp{}), while the opposite pattern emerges for male researchers. However, these differences are not statistically significant for sectors \IpPp{} and \Ipp{}. The gender differences are particularly pronounced in the outlier sectors \Ppp{} and \IPpp{}, where the deficit of female career years exceeds 40\%, and in the sector \IpPm{}, where the excess of female career years reaches approximately 20\%.

In addition to the occupation patterns, we investigate the differences in productivity and journal prestige between female and male researchers within each sector. For each sector, we quantify these gender gaps by subtracting the average productivity and journal prestige of female researchers from those of their male counterparts. As Figure~\ref{fig:2}(c) shows, female researchers exhibit lower average productivity than males across all place sectors, with the exception of the hyperprolific-perfectionist sector (\IPpp{}), where the difference is not statistically significant. The higher productivity of male researchers, however, does not translate into consistent differences in average journal impact. Indeed, Figure~\ref{fig:2}(d) reveals no statistically significant gender gaps in journal prestige within the non-outlier sectors \IpPp{} and \ImPm{}. Moreover, the average journal impact of female career years is slightly higher in sector \ImPp{} and slightly lower in sector \IpPm{} compared with male career years. Despite the substantial productivity gaps in the outlier sectors \Ipp{} and \Ppp{}, gender differences in journal prestige are not statistically significant in these sectors. Only among the small minority of researchers accessing sector \IPpp{} do female career years exhibit a lower average journal impact.

Supplementary Figure~S1 shows the analogous results of Figure~\ref{fig:1} and \ref{fig:2} when considering the SJR as a measure of journal prestige. Despite covering more researchers and disciplines (Supplementary Figure~S2), the patterns observed with SJR remain largely consistent with those previously reported for the JIF.

\subsection*{Evolution of gendered patterns in journal prestige versus productivity plane}

As with other career aspects~\cite{gingras2008effects, rorstad2015publication, sunahara2023universal}, occupation patterns in the journal prestige-productivity plane evolve as researchers advance in their careers~\cite{sunahara2021association}. It is thus natural to ask whether these changes also exhibit gendered patterns. To start addressing this question, we investigate transitions among sectors of the plane between successive career years, stratifying by gender and distinguishing between outlier and non-outlier researchers. For each possible transition, we calculate a $z$-score by subtracting the observed transition frequency from the average frequency obtained from a null model in which individual trajectories are randomly shuffled, and then dividing the result by the standard deviation of the null model outcomes (see Methods for details). Positive $z$-scores indicate transitions occurring more frequently than expected under the null model, negative $z$-scores indicate less frequent transitions, and values near zero represent transitions consistent with the null model expectation.

\begin{figure*}[ht!]
    \centering
    \includegraphics[width=0.8\linewidth]{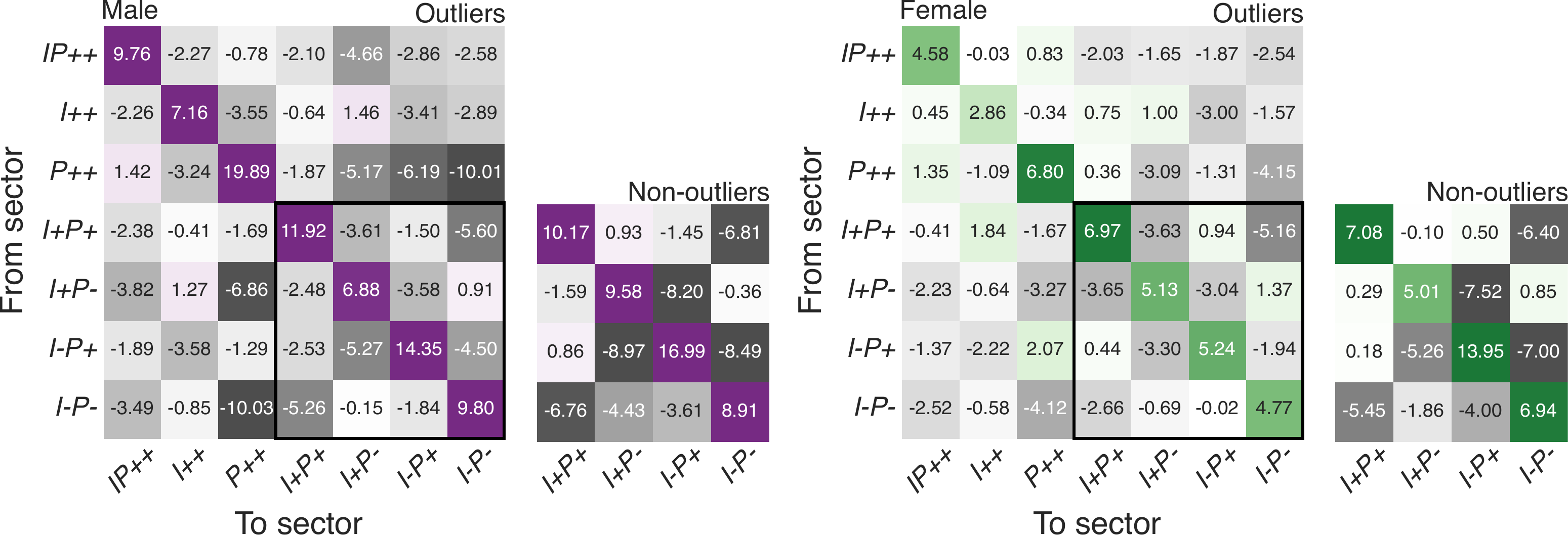}
    \caption{Gender differences in transitions between sectors of the journal prestige-productivity plane. Standardized transition frequencies between all pairs of sectors observed over the careers of male (left panel) and female (right panel) researchers, further grouped into outliers and non-outliers researchers. For each possible transition and gender, the values represent $z$-scores calculated by subtracting the observed transition frequency from the average frequency obtained via randomly shuffling researchers' trajectories and dividing by the standard deviation of the corresponding shuffled outcomes.}
    \label{fig:3}
\end{figure*}

Figure~\ref{fig:3} shows that, regardless of gender and outlier status, researchers exhibit a strong tendency to remain in the same sector between successive career years, as evidenced by the large $z$-scores along the main diagonal of the transition matrices. The secondary diagonals display exclusively negative $z$-scores, which are also among the largest in absolute value. Given the ordering of rows and columns, these antidiagonal elements correspond to transitions involving simultaneous changes in productivity and journal impact levels (\mbox{$I\!\!+\!\!P\pm$}$\ie$\mbox{$I\!\!-\!\!P\mp$}), suggesting researchers are very unlike to simultaneously shift both dimensions at consecutive career years. Moreover, the transition matrix is nearly symmetric, showing that most transitions lack a preferred temporal direction. 

Despite these overall similarities, some gender differences emerge. The most evident refers to the values in the main and secondary diagonals, which are significantly higher for male than female researchers. This indicates that male researchers are more likely to maintain their productivity and journal impact levels, and are less likely to simultaneously change both dimensions, compared to their female counterparts. Both genders show comparable likelihoods of transitioning from \Ppp{} to \IPpp{}, while transitions from \IPpp{} to \Ppp{} are rarer among male researchers. This suggests that both genders tend to access the hyperprolific-perfectionist sector from the hyperprolific sector, but once in \IPpp{}, female researchers are more likely to sustain the outlier productivity level. Female outliers also more frequently transition into the perfectionist sector from \IpPp{} and into the hyperprolific sector from \ImPp{} than male outliers. Among the non-outliers, female researchers more often transition from above- to below-average journal impact sectors (\mbox{$I\!\!+\!\!P\pm$}$\to$\mbox{$I\!\!-\!\!P\pm$}) and from \IpPm{} to \IpPp{}, whereas male researchers more commonly transition in the opposite direction, from \IpPp{} to \IpPm{}.

\begin{figure*}[ht!]
    \centering
    \includegraphics[width=0.95\linewidth]{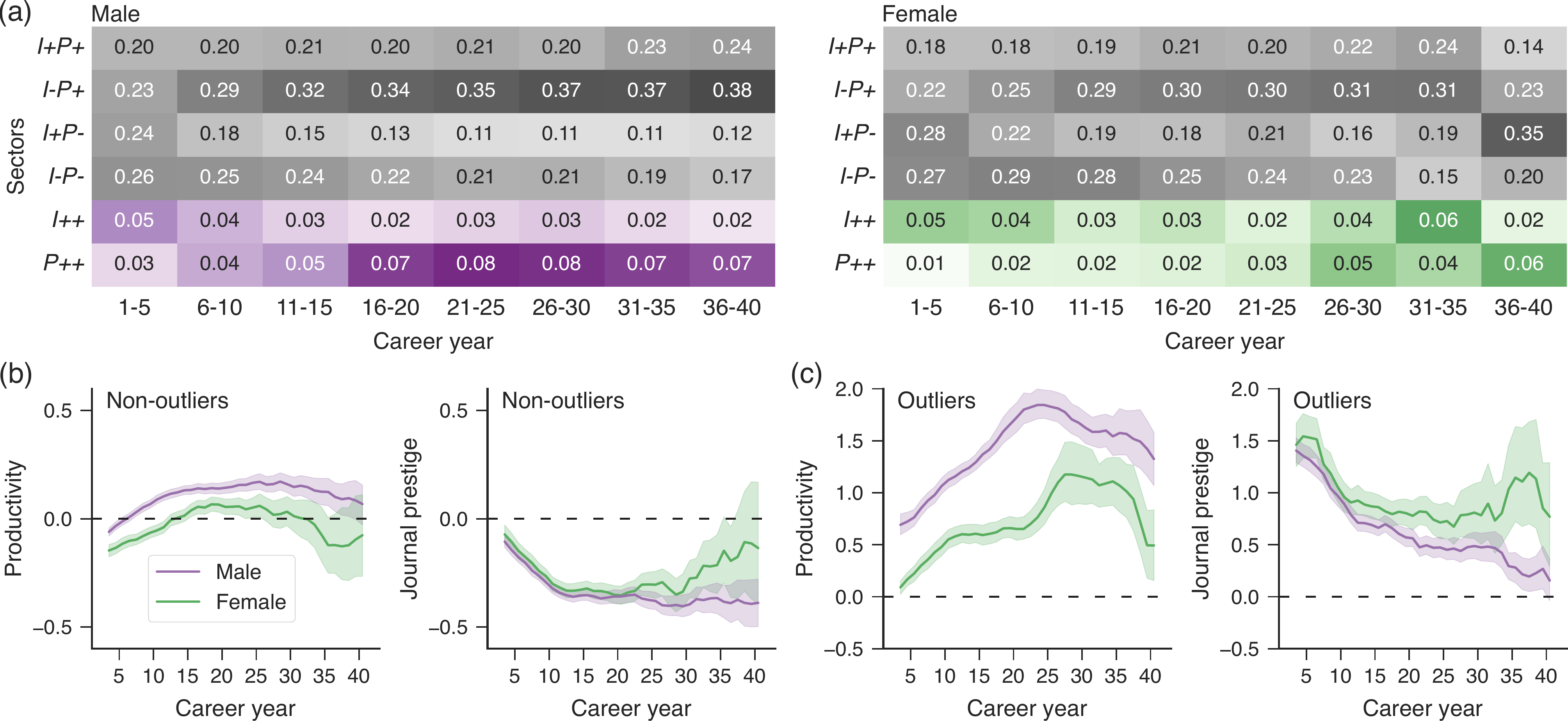}
    \caption{Evolution of gendered patterns over researchers' careers. (a) Occupation of sectors of the journal prestige-productivity plane at five-year intervals over the careers of male (left) and female (right) researchers. (b) Evolution of average productivity (left) and average journal prestige (right) over the careers of male (purple) and female (green) non-outlier researchers. (c) Evolution of average productivity (left) and average journal prestige (right) over the careers of male (purple) and female (green) outlier researchers. Shaded areas in panels (b) and (c) represent 95\% bootstrap confidence intervals, and dashed lines indicate the zero baseline.}
    \label{fig:4}
\end{figure*}

We also investigate whether the evolution of occupation across sectors of the journal prestige-productivity plane exhibits gendered patterns over the course of researchers' careers. To do so, we consider the first career year as the year following Ph.D. graduation and group the data into intervals of five career years. For each interval, we calculate the occupation fraction in each sector, stratifying by gender and further ignoring the sector \IPpp{} to ensure a minimum of 20 researchers per interval. Figure~\ref{fig:4}(a) presents the results as matrix plots, where rows refer to different plane sectors and columns stand for career-year intervals. Early career stages are more concentrated in the low-productivity sectors (\IpPm{} and \ImPm{}) and tend to shift towards the high-productivity sectors (\IpPp{} and \ImPp{}) as careers progress. However, this shift is markedly more pronounced among male researchers, who are particularly prevalent in sector \ImPp{} in later career stages, whereas female researchers consistently exhibit higher occupation fractions in sector \IpPm{} across all intervals. Both genders show an increasing trend in occupying the hyperprolific sector, although this trend is more evident among male researchers. In contrast, female researchers remain more evenly distributed within the perfectionist sector over time, whereas their male counterparts exhibit a clearer decline in occupation fraction in this sector as their careers advance.

In addition to these occupation trends, we further examine the evolution of average productivity and journal prestige across career years. Figure~\ref{fig:4}(b) depicts these average values for non-outlier researchers, stratified by gender. In the early stages of their careers, the productivity of male researchers is statistically different and higher than that of their female counterparts. As careers advance, the average productivity of male researchers reaches a plateau around the 12th career year, whereas female researchers continue to increase their productivity until approximately the 18th career year, narrowing the productivity gap (with confidence intervals nearly overlapping). After this point, the average productivity of both genders declines, with a steeper decrease observed among female researchers. A markedly different pattern emerges for average journal impact, which decreases to a plateau during the initial career years and is slightly higher among female researchers, although the difference is not statistically significant. From approximately the 20th year of their careers, the average journal impact of female researchers begins to rise, while that of male researchers remains relatively stable; yet the gap in journal prestige remains statistically non-significant. 

A similar, yet more pronounced, pattern emerges for outlier researchers. As shown in Figure~\ref{fig:4}(c), the average productivity of male outliers is considerably higher and increases at a faster pace than that of female outliers. Among male researchers, productivity peaks around the 23rd year of their career and declines thereafter. In contrast, female outliers exhibit a much later peak (approximately five years later), when the productivity gap becomes nearly statistically indistinguishable, before their productivity undergoes a sharp decline. The average journal impact of female outliers is consistently higher and statistically distinct from that of their male counterparts at several points throughout their careers, further exhibiting a u-shaped relationship with career year (peaking in both early and later stages), whereas journal prestige among male outliers decreases monotonically.

Supplementary Figures~S3 and S4 show that replacing the JIF with SJR yields very similar patterns in the transitions among plane sectors, as well as comparable trends in the occupation of plane sectors and evolution of the average productivity and journal prestige across career years.

\subsection*{Effects of productivity and career age on journal prestige at individual level}

\begin{figure*}[ht!]
    \centering
    \includegraphics[width=0.85\linewidth]{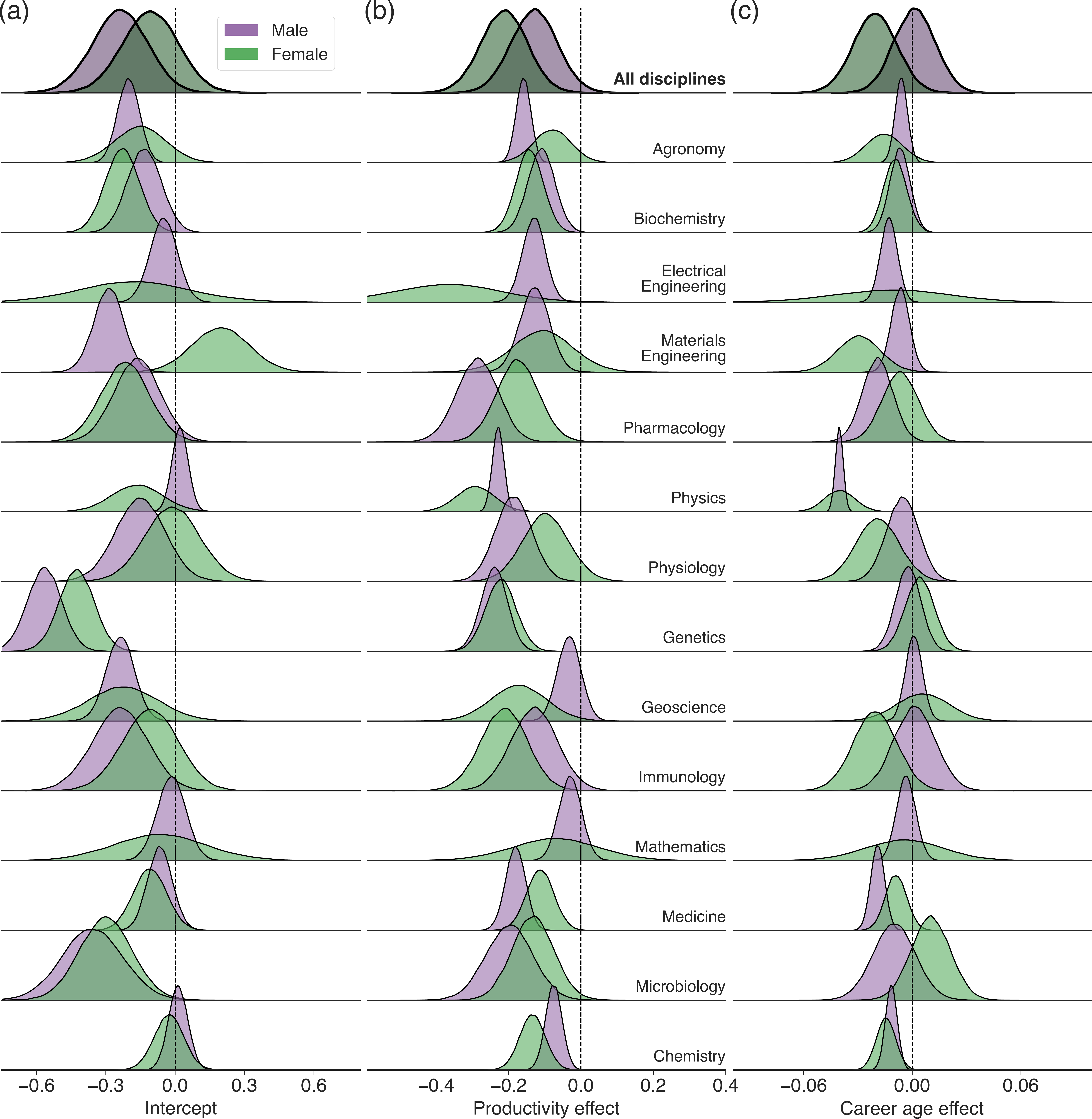}
     \caption{Role of gender in the effects of productivity and career age on average journal prestige of non-outlier researchers. Posterior probability distributions of the population-level parameters describing (a) baselines for journal prestige ($\mu_\alpha$ for males and $\mu_\alpha + \tilde{\mu}_\alpha$ for females), (b) the effect of productivity ($\mu_P$ for males and $\mu_P + \tilde{\mu}_P$ for females), and (c) the effect of career age ($\mu_A$ for males and $\mu_A + \tilde{\mu}_A$ for females). The top row presents results for all non-outlier researchers, while subsequent rows correspond to each of the 14 disciplines in the JIF dataset. Distributions for male researchers are shown in purple, and those for female researchers in green. Vertical dashed lines denote the zero baseline.}
     \label{fig:5}
\end{figure*}

Our findings thus far indicate that gender affects the relation between productivity, career age, and journal prestige when considering aggregate information of researchers' careers. To better understand the individual effects of productivity and career age on the journal prestige of male and female researchers, we employ a linear hierarchical Bayesian model. As detailed in the Methods section, we model the average journal impact ($I_j$) of a researcher $j$ as a linear function of their productivity ($P_j$) and career age ($A_j$), allowing the coefficients to depend on gender. Specifically, we consider $I_j \sim (\alpha_j+\tilde{\alpha}_j G_j) + (\beta_j + \tilde{\beta}_j G_j) P_j + (\gamma_j + \tilde{\gamma}_j G_j) A_j$, where $G_j$ is a gender indicator variable ($G_j = 1$ for female and $G_j=0$ otherwise), $\beta_j$ and $\tilde{\beta}_j$ represent the effect of productivity, $\gamma_j$ and $\tilde{\gamma}_j$ quantify the effect of career age, and the intercepts $\alpha_j$ and $\tilde{\alpha}_j$ refer to the expected baseline journal prestige at beginning of researchers career with productivity equal to the average in their research area and year. The coefficients $\alpha_j$, $\tilde{\alpha}_j$, $\beta_j$, $\tilde{\beta}_j$, $\gamma_j$, and $\tilde{\gamma}_j$ are assumed to be drawn from normal distributions with means $\mu_\alpha$, $\tilde{\mu}_\alpha$, $\mu_P$, $\tilde{\mu}_P$, $\mu_A$, and $\tilde{\mu}_A$, respectively. The Bayesian inference treats each career individually and yields the posterior distributions of these parameters, allowing us to quantify the population-level baselines for journal prestige ($\mu_\alpha$ for males and $\mu_\alpha + \tilde{\mu}_\alpha$ for females) as well as the effects of productivity ($\mu_P$ for males and $\mu_P + \tilde{\mu}_P$ for females) and career age ($\mu_A$ for males and $\mu_A + \tilde{\mu}_A$ for females). We apply this model to all non-outlier researchers with careers spanning more than 5 years, further grouping individuals by discipline to investigate whether gender effects vary across disciplines.

Figures~\ref{fig:5}(a), \ref{fig:5}(b), and \ref{fig:5}(c) show the posterior distributions for the baseline journal prestige and the estimated effects of productivity and career age on journal prestige, respectively. When considering all researchers, the baseline journal prestige for male researchers is slightly more shifted toward negative values than that of their female counterparts, suggesting that early-career male researchers with productivity close to the average of their disciplines in a given year tend to publish in journals of lower average prestige than comparable female researchers. This pattern, however, varies across disciplines. In some areas, such as Pharmacology and Medicine,  the posterior distributions for both genders are nearly identical, while in others, such as Physics, the distribution is more shifted toward negative values for female researchers. Similarly, although productivity is negatively associated with journal impact -- with a slightly stronger effect among female researchers --  gender differences are discipline-specific. For instance, in Pharmacology, the average journal prestige of female researchers is less negatively affected by productivity than their male counterparts, whereas the opposite holds in other disciplines, such as Geoscience and Immunology. The effect of career age on journal prestige is significantly smaller than that of productivity, with aggregated results across all disciplines indicating that the journal prestige of female researchers is negatively impacted by career age, whereas the effect among male researchers centers around zero; yet these gender differences exhibit substantial variation across disciplines. Supplementary Figure~S5 shows that replacing JIF with SJR yields qualitatively similar results, corroborating the complexity of gender effects and emphasizing the discipline-dependent nature of the association between journal prestige, productivity, and career age.

\section*{Discussion}

Our findings thus reveal distinct patterns in the association between productivity and journal prestige across the careers of male and female researchers belonging to the Brazilian scientific elite. By mapping individual career years onto a standardized journal prestige-productivity plane, we find that male researchers are disproportionately represented in productivity-oriented sectors of the plane, particularly in the hyperprolific region. Conversely, female researchers are more prevalent in sectors that do not prioritize publication quantity, especially within the non-outlier sector that favors average journal prestige over productivity. Furthermore, although male researchers consistently exhibit higher productivity across all sectors of the plane, gender differences in average journal prestige are statistically non-significant in most sectors, with female researchers achieving slightly higher prestige in one productivity-focused sector and slightly lower in another where impact is prioritized. Among outlier individuals, we observe that each additional hyperprolific year more strongly reduces the likelihood of simultaneously achieving perfectionist performance for female researchers than for their male counterparts.

We further investigate temporal aspects underlying the occupation of the journal prestige-productivity plane throughout academic careers. This analysis demonstrates that male researchers exhibit greater persistence in maintaining their productivity and journal prestige levels, showing notably greater aversion to simultaneously changing both metrics between successive career years when compared to female researchers. Researchers of both genders are more prevalent in low-productivity sectors during the early stages of their careers, yet the shift towards high-productivity sectors is more pronounced among male researchers. As careers advance, male researchers also reduce their presence in the perfectionist sector, while female researchers maintain a more stable representation in this sector. Moreover, male researchers consistently achieve higher productivity than their female peers at all career stages; still, the average journal prestige levels are largely comparable between genders. The productivity gap is particularly pronounced among outliers, but this disparity is also not reflected in the average journal prestige. Instead, the average journal prestige of female outliers exceeds that of their male peers, especially in later career stages. This pattern reflects gender trajectories that diverge in terms of average journal prestige, which monotonically decreases with career progression for male outlier researchers while follows a u-shaped trajectory for female peers.

In addition to examining aggregated patterns, we use a Bayesian hierarchical approach to probe gendered patterns in the joint effect of productivity and career age on average journal prestige at the individual level. Our results indicate that productivity and career age negatively impact the average journal prestige of researchers from both genders, with slightly stronger effects observed among female researchers. The baseline journal prestige -- corresponding to the expected prestige at career onset for researchers whose productivity matches the average of their discipline in a given year -- is also marginally higher for female researchers. Yet, discipline-specific analyses reveal heterogeneous gender effects, with some disciplines exhibiting negligible or even reversed gender differences, while others display more pronounced disparities in how productivity and career age relate to journal impact.

Taken together, beyond the numerical overrepresentation of male scholars, our research suggests that female and male researchers navigate the productivity-prestige landscape differently, yet ultimately attain comparable journal impact despite persistent productivity gaps. These differences likely reflect a combination of individual choices and structural factors, such as unequal access to resources, different career incentives, varying risk tolerances, and other systemic constraints. Life and family aspects -- most notably the transition to parenthood~\cite{fan2024parent, morgan2021parenthood, cech2019parents} and the division of domestic labor -- may also contribute to the observed gender differences. While our dataset precludes a direct analysis of parenthood, the literature consistently identifies a ``motherhood penalty~\cite{fan2024parent},'' whereby female researchers often bear a disproportionate share of caregiving and domestic responsibilities compared to their male peers. This external pressure imposes a substantial time tax that may effectively limit the capacity for hyper-prolific publication strategies more frequently observed among male researchers. Consequently, the tendency for female scholars to occupy regions of the productivity-prestige plane that prioritize average journal impact over volume may reflect a necessary optimization under time constraints, rather than a simple difference in academic preferences. Moreover, our results reinforce the idea that productivity, career age, and impact are not tightly coupled; instead, their interrelations at the individual level proved to be discipline-specific and complex in terms of gender, highlighting that different productivity strategies can yield similar outcomes in terms of average journal prestige.

Our work is however not without its limitations. Although our findings reveal gendered patterns from a major research ecosystem outside North America and Europe, they may not be generalizable to other countries due to differences in research dynamics, funding structures, and national scientific policies. Moreover, our dataset focuses on a relatively homogeneous group of elite researchers working predominantly in science, technology, engineering, and mathematics (STEM) disciplines. The dominance of STEM fields in our dataset reflects a specific publishing culture, and while this focus was necessary to produce robust standardized measures, we acknowledge that it may not align with publication norms in other scientific domains, such as the Humanities or Social Sciences, which exhibit distinct patterns of publication and journal venue selection. This leaves open the question of whether the observed disparities and similarities extend to the entire Brazilian research workforce, since non-elite cohorts or researchers outside of STEM may exhibit different gendered patterns and career dynamics. Apart from career age, we do not directly account for other mechanisms, such as funding disparities or collaboration networks, that may differentially affect the career trajectories of male and female researchers. Other factors, such as race, socioeconomic status, and institutional affiliation, may also influence the patterns observed. In addition, although journal-based metrics correlate with citations and carry information about academic performance, they do not capture the natural variability among articles published by the same journal or reflect the quality and innovativeness of the individual publications~\cite{liu2023interdisciplinary, zhang2024composition, liu2024novelty}. Future work may then focus on addressing these limitations by integrating large-scale bibliometric datasets with survey information to capture additional dimensions that may intersect with our binary gender classification. Despite these limitations, our findings reinforce the need to revisit dominant paradigms of research evaluation, which may disproportionately reward productivity-focused trajectories and undervalue impact-oriented or non-linear career paths more commonly pursued by female researchers -- recognizing and rewarding these alternative trajectories contributes to a more equitable and diverse scientific community, which, beyond societal implications, also yields economic and scientific relevant benefits.

\section*{Methods}

\subsection*{Dataset description}

Our dataset comprises the publication records of Brazilian researchers holding the Research Productivity Fellowship (\textit{Bolsa de Produtividade em Pesquisa}) as of May 2017. This nationally prestigious fellowship, awarded by the Brazilian National Council for Scientific and Technological Development (CNPq -- Conselho Nacional de Desenvolvimento Cient\'ifico e Tecnol\'ogico) since the 1970s, recognises scholars producing high‑quality research. Originally introduced in Ref.~\cite{sunahara2021association}, the dataset initially included 14,487 researchers from 88 disciplines, for whom the complete publication records were retrieved from the Lattes Platform (\textit{Plataforma Lattes})~\cite{cnpq1999lattes} -- Brazil's official academic curriculum database. The prestige of these articles was estimated from their publication venues using Clarivate's Journal Impact Factor (JIF) and Scopus's SCImago Journal Rank (SJR), which spanned the periods 1997-2015 and 1999-2015, respectively. Annual productivity (number of articles per year) and the respective average journal prestige were calculated by aligning each researcher's curriculum data with the time-varying JIF and SJR values. The combined data were then grouped by discipline and year, retaining only researchers from disciplines with at least 50 researchers with published articles in each year (ensuring robust discipline‑level averages for defining the standardized metrics). The final JIF-based dataset comprises 6,028 researchers across 14 disciplines, whereas the SJR-based dataset includes 8,465 researchers from 25 disciplines. Both datasets are primarily composed of researchers working in science, technology, engineering, and mathematics (STEM) disciplines (Supplementary Figure~2).

Gender information is original to this work. The gender of each researcher (male or female) was determined using name-based gender prevalence data from the Brazilian Census~\cite{ibge2010nomes}. When this approach proved inconclusive, we manually assigned gender by examining publicly available profile pictures on the Lattes Platform or, when necessary, by contacting the researchers' institutions.

\subsection*{Standardized measures of research productivity and journal impact}

Following Ref.~\cite{sunahara2021association}, and to compare the productivity and average journal prestige over the careers of researchers from different disciplines, we used a standardized approach to define a normalized measure of productivity relative to discipline and year, as well as a normalized measure of average journal impact that, in addition to discipline and year, is further defined relative to researchers' productivity level. Letting $p^k_j(y)$ denote the number of papers published by researcher $j$ from discipline $k$ in year $y$, the normalized measure of productivity is defined as 
\begin{equation*}
    P^k_j(y) = \dfrac{p^k_j(y) - \mathbb{E}[p^k(y)]}{\mathbb{S}[p^k(y)]},
\end{equation*}
where $\mathbb{E}[p^k (y)]$ and $\mathbb{S}[p^k (y)]$ correspond to the average and standard deviation of the productivity of all researchers from discipline $k$ in year $y$, respectively. This definition thus accounts for discipline-specific publication practices (by comparing a researcher's productivity to that of their peers from the same discipline) as well as inflationary trends (related to the increasing number of papers published annually)~\cite{sinatra2016quantifying, sunahara2021association}.

In its turn, the normalized measure of journal prestige is defined as 
\begin{equation*}
    I^k _j(y) = \dfrac{i^k_j(y) - \mathbb{E}[i^k_{\text{rnd}} (y,p^k(y))]}{\mathbb{S}[i^k_{\text{rnd}}(y,p^k(y))]},
\end{equation*}
where $i^k_j(y)$ is average journal prestige of the publications by researcher $j$ from discipline $k$ in year $y$, $i^k_{\text{rnd}}(y,p)$ corresponds to the average journal prestige of a random sample of $p^k(y)$ publications from discipline $k$ in year $y$, and $\mathbb{E}[i^k_{\text{rnd}}(y,p^k(y))]$ and $\mathbb{S}[i^k_{\text{rnd}}(y,p^k(y))]$ represent the average and standard deviation of $i^k _{\text{rnd}}(y,p^k(y))$, respectively, estimated over 1,000 independent realizations of the sampling process. Thus, in addition to being relative to discipline and year, this measure further accounts for the fact that, due to statistical fluctuations, high-productivity researchers tend to show lower variability in average journal prestige, while those with fewer publications display greater variability. This formulation was inspired by an analogous index introduced by Antonoyiannakis for comparing the impact factor of journals of different sizes~\cite{antonoyiannakis2018impact, antonoyiannakis2020impact}.

In addition, to account for the asymmetric distribution of raw productivity and average journal prestige values, as well as the presence of researchers with exceptionally high values in these metrics, we replaced the conventional definitions of mean and standard deviation with their corresponding Huber robust estimators~\cite{huber2004robust}, as implemented in the Python package \texttt{statsmodels}~\cite{seabold2010statsmodels}.

\subsection*{Normalized transitions among sectors of journal prestige versus productivity plane}

We evaluated the transitions among sectors of the journal prestige-productivity plane using a standard score measure relative to a null model based on shuffled versions of researchers' careers. Specifically, for each gender, we calculated 
\begin{equation*}
    Z_{ij} = \dfrac{M_{ij} - \mathbb{E}[M^{\text{rnd}}_{ij}]}{\mathbb{S}[M^{\text{rnd}}_{ij}]},
\end{equation*}
where $M_{ij}$ denotes the number of observed transitions from sector $i$ to sector $j$, $\mathbb{E}[M^{\text{rnd}}_{ij}]$ and $\mathbb{S}[M^{\text{rnd}}_{ij}]$ represent the mean and standard deviation, respectively, of the number of transitions obtained from 1,000 independent realizations of the null model generated by randomly shuffling the career metrics of researchers.

\subsection*{Effects of hyperprolific years on the likelihood of performing as a perfectionist}

We applied logistic regression to evaluate gender differences in the association between being an outlier in productivity and the probability $\Pi_{\text{perfectionist}}$ of also being an outlier in journal prestige, via the model
\begin{equation*}
    \Pi_{\text{perfectionist}} = \frac{e^{a_0 + a_1 Y_P}}{1 - e^{a_0 + a_1 Y_P}},
\end{equation*}
where $Y_P$ denotes the number of outlier career years in productivity, and $a_0$ and $a_1$ are the intercept and the logistic regression coefficient, respectively. Negative values of $a_1$ indicate that an increase in $Y_P$ reduces the likelihood of performing as a perfectionist, whereas positive values indicate an increased likelihood. The model parameters were estimated using the logistic regression implementation in \texttt{statsmodels}~\cite{seabold2010statsmodels}, separately for male and female outlier researchers, and the resulting $a_1$ values were statistically significant ($\text{p-value} < 0.001$). For male researchers, we obtained $a_0 = 0.99\pm0.07$ (95\% confidence interval: $[0.85,1.13]$) and $a_1 = -0.24\pm0.02$ ($[-0.29;-0.19]$); for female researchers $a_0 = 2.06\pm0.15$ ($[1.76,2.36]$) and $a_1 = -0.48\pm0.07$ ($[-0.62,-0.35]$). We note that the coefficients in both regressions differ significantly between the two genders. These parameters were used to construct the curves shown in Figure~\ref{fig:1}(c). 

\subsection*{Bayesian hierarchical modeling of the association between productivity and career age on journal prestige}

We applied a Bayesian hierarchical model to assess the impact of productivity and career age on the average journal prestige across the careers of all non-outlier researchers in our dataset. Specifically, we considered that observations of average journal impact $I_j$, productivity $P_j$, career age $A_j$, and gender $G_j$ ($G_j=1$ for female and $G_j=0$ otherwise) are nested within a researcher $j$, and modeled the association between $I_j$ and the other variables via a linear relationship
\begin{equation*}
    I_j \sim {N}\left[\!(\alpha_j+\tilde{\alpha}_j G_j) + (\beta_j + \tilde{\beta}_j G_j) P_j + (\gamma_j + \tilde{\gamma}_j G_j) A_j, \varepsilon\right],
\end{equation*}
where ${N}[\mu,\sigma]$ denotes a normal distribution with mean $\mu$ and standard deviation $\sigma$. Parameters $\alpha_j$ and $\tilde{\alpha}_j$ represent the model intercept, $\beta_j$ and $\tilde{\beta}_j$ quantify the effect of productivity, $\gamma_j$ and $\tilde{\gamma}_j$ quantify the effect of career age, and $\varepsilon$ accounts for unobserved determinants of $I_j$. In addition, in our hierarchical Bayesian formulation, the model coefficients are considered random variables drawn from the following normal distributions:
\begin{equation*}
    \begin{aligned}
        \alpha_j &\sim {N}[\mu_{\alpha}, \sigma_{\alpha}],  &\tilde{\alpha}_j &\sim {N}[\tilde{\mu}_{\alpha}, \tilde{\sigma}_{\alpha}] \\
        \beta_j  &\sim {N}[\mu_{P}, \sigma_{P}],            &\tilde{\beta}_j  &\sim {N}[\tilde{\mu}_{P}, \tilde{\sigma}_{P}] \\
        \gamma_j &\sim {N}[\mu_{A}, \sigma_{A}],            &\tilde{\gamma}_j &\sim {N}[\tilde{\mu}_{A}, \tilde{\sigma}_{A}]
    \end{aligned}\,,
\end{equation*}
whose parameters (or hyperparameters) are also random variables (drawn from the hyperpriors distributions). This model thus allows researchers to have their own parameters, while avoiding overfitting by ensuring that individual parameters are centered around global averages. 

The parameters $\mu_P$ and $\tilde{\mu}_{P}$ quantify the population-level effects of productivity on average journal impact, with $\mu_P$ representing the effect for male researchers and $\tilde{\mu}_{P}$ representing the difference in effect for female researchers; thus, $\mu_P+\tilde{\mu}_{P}$ corresponds to the effect for female researchers. Similarly, $\mu_A$ and $\tilde{\mu}_{A}$ quantify the population-level effects of career age on average journal impact, where $\mu_A$ represents the effect for male researchers and $\mu_A + \tilde{\mu}_{A}$ for female researchers. The intercept parameters, $\mu_{\alpha}$ and $\mu_{\alpha}+\tilde{\mu}_{\alpha}$, represent the expected journal prestige for male and female researchers, respectively, at the start of their careers, assuming a productivity equal to the average of their disciplines in a given year. The Bayesian inference process consists of estimating the posterior probability distributions of intercepts and linear coefficients, both at the individual and population levels. 

Figure~\ref{fig:5} and Supplementary Figure~5 depict the posterior distributions for $\mu_{\alpha}$ and $\mu_{\alpha}+\tilde{\mu}_{\alpha}$ (population-level intercepts for male and female researchers, respectively), $\mu_P$ and $\mu_P + \tilde{\mu}_{P}$ (population-level effect of productivity on journal prestige for male and female researchers), and $\mu_{A}$ and $\mu_{\alpha}+\tilde{\mu}_{A}$ (population-level effect of career age on journal prestige for male and female researchers). We estimated these distributions both across all researchers in each dataset as well as separately for each of the 14 disciplines in the JIF dataset and the 25 disciplines in the SJR dataset.

To avoid introducing bias into the posterior estimates, we adopted the following non-informative prior and hyperprior distributions~\cite{gelman2006priordistributions}:
\begin{equation*}
    \begin{aligned}
        \varepsilon &\sim {U}(0,10^2) \\
        \mu_{\alpha} &\sim {N}(0, 10^5), & \tilde{\mu}_{\alpha} &\sim  {N}(0, 10^5)\\
        \mu_{P} &\sim {N}(0, 10^5), & \tilde{\mu}_{P} &\sim  {N}(0, 10^5)\\
        \mu_{A} &\sim {N}(0, 10^5), & \tilde{\mu}_{A} &\sim  {N}(0, 10^5)\\
        \sigma_{\alpha} &\sim  \text{Inv-}\Gamma(10^{-3}, 1), & \tilde{\sigma}_{\alpha} &\sim   \text{Inv-}\Gamma(10^{-3}, 1)\\
        \sigma_{P} &\sim  \text{Inv-}\Gamma(10^{-3}, 1), & \tilde{\sigma}_{P} &\sim   \text{Inv-}\Gamma(10^{-3}, 1)\\
        \sigma_{A} &\sim  \text{Inv-}\Gamma(10^{-3}, 1), & \tilde{\sigma}_{A} &\sim   \text{Inv-}\Gamma(10^{-3}, 1)\\
    \end{aligned}\,,
\end{equation*}
where ${U}(x_{\text{min}},x_{\text{max}})$ denotes a uniform distribution between $x_{\text{min}}$ and $x_{\text{max}}$, and $\text{Inv-}\Gamma(a,b)$ represents an inverse-gamma distribution with shape parameter $a$ and scale parameter $b$. Bayesian inference was implemented using the PyMC3 framework~\cite{pymc3} via the gradient-based Hamiltonian Monte Carlo No-U-Turn-Sampler method to draw samples from the posterior distributions. We ran eight parallel chains with 10,000 iterations (discarding the first 5,000 as burn-in) for each regression, and assessed convergence using the Gelman-Rubin statistic ($\hat{R}$), which was close to one across all regressions, indicating the well-mixing and convergence of the chains.

\section*{Funding}
We acknowledge the support of the Coordena\c{c}\~ao de Aperfei\c{c}oamento de Pessoal de N\'ivel Superior, the Conselho Nacional de Desenvolvimento Cient\'ifico e Tecnol\'ogico (CNPq -- Grants 303533/2021-8), and the Slovenian Research Agency (Grants J1-2457 and P1-0403).

\section*{Author contributions statement}
V.H.R., A.S.S., G.S., M.P., and H.V.R. designed research, performed research, analyzed data, and wrote the paper.
 
\section*{Data availability}
The data are available from the corresponding authors on reasonable request.

\bibliography{references}
\clearpage
\includepdf[pages=1-6,pagecommand={\thispagestyle{empty}}]{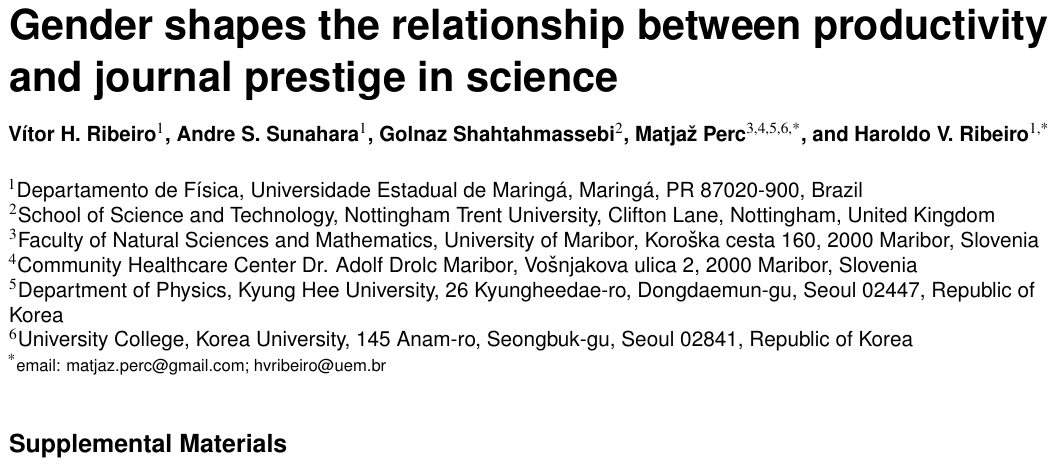}

\end{document}